\documentstyle[aps,pre,epsfig,floats]{revtex}
\begin{document}
\draft \twocolumn[\hsize\textwidth\columnwidth\hsize\csname
@twocolumnfalse\endcsname

\title{A new mechanism of
hypersensitive transport in tilted sharp ratchets\\
induced by noise flatness}
\author{
  Romi Mankin,${}^1$\ %
  Astrid Haljas,${}^{2,1}$\ %
  Risto Tammelo,${}^{2,*}$\ %
  \ and \ Dmitri Martila${}^2$%
  }
 \address{${}^1$Department of Natural Sciences,
 Tallinn Pedagogical
  University, 25 Narva Road, 10120 Tallinn, Estonia\\
  ${}^2$Institute of Theoretical ~Physics, Tartu University,
  4 T\"ahe
  Street, 51010 Tartu, Estonia}

\date{\today}
\maketitle

\begin{abstract}
The noise-flatness-induced hypersensitive transport of overdamped
Brownian particles in a tilted ratchet system driven by
multiplicative nonequilibrium three-level Markovian noise and
additive white noise is considered. At low temperatures the
enhancement of current is very sensitive to the applied small
static tilting force. It is established that the enhancement of
mobility depends non-monotonically on the parameters (flatness,
correlation time) of multiplicative noise. The optimal values of
noise parameters maximizing the mobility are found.
\end{abstract}

 ]

Recently, noise-induced hypersensitivity to small time-dependent
signals in nonlinear systems with multiplicative noise has been
the topic of a number of physical investigations
\cite{gin98,gin01,ben,ger00}. A motivation in this field has come
from numerical, analytical, and experimental studies of a
nonlinear Kramers oscillator with multiplicative white noise
\cite{gin98,ger00}. Under the effect of intense multiplicative
noise, the system is able to amplify an ultrasmall deterministic
ac signal (of the order of, e.g., $10^{-20}$) up to the value of
the order of unity \cite{gin98}. Afterwards, a related phenomenon
such as noise-induced hypersensitive transport was found in some
other systems with multiplicative dichotomous noise
\cite{gin01,ben}. Noise-induced hypersensitive transport was also
established in a phase model, i.e., $d\varphi/dt=a-b\sin\varphi$,
with a strong symmetric multiplicative colored noise. It was shown
that in such a system a macroscopic flux (current) of matter
appears  under the effect of ultrasmall dc driving \cite{gin01}.
It is important to notice that the physical mechanism underlying
the phenomenon of hypersensitive transport presented in Refs.
\cite{gin01,ben} is based on the assumption that the periodic
potential is smooth. It is easy to see that in the case of a
periodic sharp potential the above mechanism cannot bring forth
any hypersensitive transport.

Theoretical investigations  \cite{doe94,bie,els,berg,mankin}
indicate that noise-induced nonequilibrium effects are sensitive
to noise flatness, which is defined as the ratio of the fourth
moment to the square of the second moment of the noise process.
Although its significance is obvious, the role of the flatness of
fluctuations has not been researched to any significant degree
to-date. In the present paper we assume the multiplicative noise
to be a zero-mean trichotomous Markovian stochastic process
\cite{man}. It is remarkable that for trichotomous noises the
flatness parameter $\varphi$, contrary to the cases of the
Gaussian colored noise $(\varphi=3)$ and symmetric dichotomous
noise $(\varphi=1)$, can have any value from 1 to $\infty$. The
flatness as an extra degree of freedom (in comparison with
dichotomous noise) can prove useful when modelling actual
fluctuations, e.g., thermal transitions between three
configurations or states. This is the reason why we choose in the
phase space of possible non-equilibrium models the trichotomous
noise. Although both dichotomous and trichotomous processes may be
too rough approximations of the actual non-equilibrium
fluctuations, the latter is more flexible, including all cases of
dichotomous processes and, as such, revealing the essence of its
peculiarities. A further virtue of the models with trichotomous
noise is that they constitute a case admitting exact analytical
solutions for some nonlinear stochastic problems, such as
colored-noise-induced transitions \cite{man} and reversals of
noise-induced flow \cite{mankin}.

The main purpose of this paper is to establish a new mechanism of
hypersensitive transport, demonstrating that the flatness of
multiplicative noise can generate hypersensitive response to the
small external static force in a tilted sharp ratchet system. We
will show that in the region of hypersensitive response the value
of mobility can be controlled by means of thermal noise. For low
temperatures, we find that the mobility exhibits resonant behavior
at intermediate values of the parameters of the multiplicative
noise (flatness, correlation time).

We consider an overdamped multinoise tilted ratchet, where
particles move in a one-dimensional spatially periodic potential
of the form $V(x,t)=V(x)Z(t)$, where $Z(t)$ is a trichotomous
process \cite{man} and $V(x)$ is a piecewise linear function,
which has one maximum per period. The additional force consists
of thermal noise with temperature $D$, and an external static
force $F$. The system is described by the dimensionless Langevin
equation
\begin{equation}
{dX\over dt}=Z(t)h(X)+F+\xi(t),\;\;\;h(x)\equiv- \frac{dV(x)}{dx}
\quad , \label{langevin}
\end{equation}
where $V(x)=\tilde{V}(\tilde{x})/\tilde{V}_0,\,
\tilde{V}(\tilde{x})$ is a spatially periodic function with period
$\tilde{L}$ and $\tilde{V}_0=\tilde{V}_{max}- \tilde{V}_{min}$.
The usual dimensionalized physical variables are indicated by
tildes and the space and time coordinates read
$X=\tilde{X}/\tilde{L}$ and $t=\tilde{t}\tilde{V}_0/
\kappa\tilde{L}^2$ with $\kappa$ being the friction coefficient;
$\tilde{F}=\tilde{V}_0F/\tilde{L}$ is a constant external force.
The thermal noise satisfies $\langle{\xi(t)}\rangle=0$ and
$\langle{\xi(t_1)\xi(t_2)}\rangle=2D\delta(t_1-t_2)$. Regarding
the random function $Z(t)$, we assume  it to be a zero-mean
trichotomous Markovian stochastic process \cite{man} which
consists of jumps among three values $z=\{1, 0, -1\}$. The jumps
follow in time according to a Poisson process, while the values
occur with the stationary probabilities $P_s(1)=P_s(-1)=q$ and
$P_s(0)=1-2q$. In a stationary state, the fluctuation process
statisfies $\langle{Z(t)}\rangle=0$ and $\langle{Z(t+\tau})Z(t)
\rangle=2q\exp(-\nu\tau)$, where the switching rate $\nu$ is the
reciprocal of the noise correlation time $\tau_{c}=1/\nu$. The
trichotomous process is a special case of the kangaroo process
\cite{doe94} with a flatness parameter
$\varphi=\langle{Z^4(t)}\rangle/\langle{Z(t)}\rangle^2=1/(2q)$. At
large flatnesses our trichotomous noise essentially coincides with
the three-level noise used by Bier \cite{bie} and Elston and
Doering \cite{els}.

The master equation corresponding to Eq.~(\ref{langevin}) reads
\begin{equation}
{\partial \over\partial t}P_n(x,t) = -{\partial \over\partial
x}[\Gamma_n P_n(x,t)] +\sum_mU_{nm}P_m(x,t) \quad, \label{master}
\end{equation}
where $\Gamma_n\!=\!z_n\,h(x)\!+\!F\!-\!D\,\partial_x$ and
$P_n(x,t)$ is the prob\-a\-bil\-i\-ty density for the combined
process $(x,z_n,t)$, $n,m=1,2,3,$ $z_1=1, \:z_2=0,\:z_3=-1$, and
$U_{ik}=\nu[q+(1-3q)\delta_{i2}-\delta_{ik}]$. The stationary
current $J=\sum_nj_n(x)$ is then evaluated via the current
densities $j_n(x)=(z_nh(x)+F-D\,\partial_x)P_n^s(x)$, where
$P_n^s(x)$ is the stationary probability density in the state
$(x,z_n)$. To calculate the stationary probability density in the
$x$ space, $P(x)=\sum_nP_n^s(x)$, and the stationary current,
$J=\textrm{const}$, six conditions are imposed on the solutions
of Eq.~(\ref{master}), namely, the conditions of periodicity
$P_n^s(x)=P_n^s(x+1),\,n=1,2,3,$ and normalization of $P_n^s(x)$
over the period interval $L=1$ of the ratchet potential
$Z(t)V(x)$, which read $\,\int_0^1P_1^s(x)dx=\int_0^1P_3^s(x)dx=q$
and $\;\int_0^1P_2^s(x)dx=1-2q$.

To derive an exact formula for $J$, we assume that the potential
$Z(t)V(x)=Z(t)V(x-1)$ in Eq.~(\ref{langevin}) is piecewise linear
(sawtoothlike) and its asymmetry is determined by a parameter
$d\in(0,1)$, with $V(x)$ being symmetric when $d=1/2$. A schematic
representation of the three configurations assumed by the "net
potentials" $V_n(x)=z_nV(x)-Fx$ associated with the right hand
side of Eq.~(\ref{langevin}) is shown in Fig.~1. Regarding the
symmetry of the dynamic system (\ref{langevin}), we notice that
$J(-F)=-J(F)$ and $J(F,d)=J(F,1-d)$. Thus we may confine ourselves
to the case $d\le1/2$ and $F\ge0$. Obviously, for $F=0$, the
system is effectively isotropic and no current can occur. In the
case of zero temperature the both noise levels $z_{n=1,3}=\pm1$ in
Eq.~(\ref{langevin}), where $F\le\min\{1/d,1/(1-d)\}$, give zero
flux. However, if one allows switching between the three dynamic
laws $V_n(x)$, $n=1,2,3,$ the resulting motion will have a net
flux which can be much greater that the flux by the dynamic law
$V_2=-Fx$. If the rate of reaching the minimal energy in each well
considerably exceeds the switching rate $\nu$, the leading part of
the net flux is achieved in the following way: a particle locked
in the potential minimum 1 switches to point 2, then slowly moves
to point 3, switches to point 4 (or to 5 with equal probability),
and rapidly slides down to point 6 (or from 5 back to 1), etc (see
Fig.~1 and cf Ref. \cite{berd}).
\begin{figure} \centerline{
\psfig{file=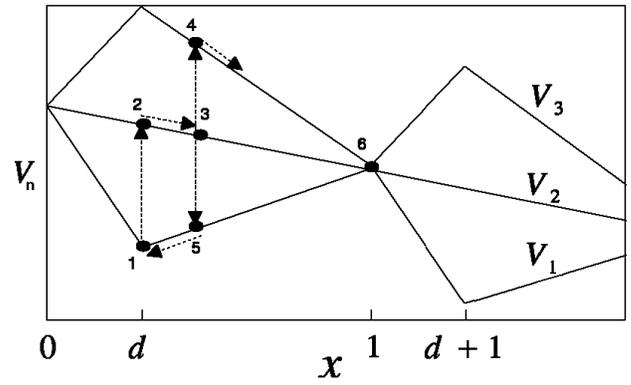,width=83mm}}\caption{The mechanism of
hypersensitive transport. The lines depict the net potentials
$V_n(x)=z_nV(x)-Fx$ with $z_1=1$, $z_2=0$, and $z_3=-1$. A
particle cannot move of its own accord along the potentials $V_1$
and $V_3$. However, if one allows switching between the potentials
$V_n$, $n=1,2,3$, the particle will move downhill along the
trajectory 1:2:3:4:6.}
\end{figure}
In this case hypersensitive transport is possible and can be
intuitively understood. The described scheme is valid only in the
absence of thermal noise. Otherwise, a particle is able to pass by
a thermally activated escape across the potential barriers in both
directions. However, it predominantly moves to the right and
hypersensitive transport still occurs (at least at sufficiently
low temperatures). As the "force" $h(x)=-dV(x)/dx$ is piecewisely
constant, $h(x)=h_1=1/d$ for $x\in(0,d)\:( \textrm{mod1})$ and
$h(x)=h_2=-1/(1-d)$ for $x\in(d,1)\:( \textrm{mod1})$,
Eq.~(\ref{master}) splits up into two linear differential
equations with constant coefficients for the two vector functions
${\bf{P_i^s}(x)}=(P_{1i}^s,P_{2i}^s, P_{3i}^s)$ $(i=1,2)$ defined
on the intervals $(0,d)$ and $(d,1)$, respectively. The solution
reads
\begin{equation}
P_{ni}^s(x)=JA_n+\sum_{k=1}^5C_{ik}A_{nik}e^{\lambda_{ik}x}\quad ,
\label{solution}
\end{equation}
where $C_{ik}$ are constants of integration, the constants $A_n$
and $A_{nik}$ are given by $A_1=A_3=qJ/F$, $A_2=(1-2q)J/F$,
$A_{nik}=(D\lambda_{ik}-F)[D\lambda^2_{ik}-(F-z_nh_i)
\lambda_{ik}-\nu]$ for $n=1,3$, $A_{2ik}=2h^2_i
\lambda_{ik}-(A_{1ik}+A_{3ik})$, and $\{\lambda_{ik},k=1,
\ldots,5\}$ is the set of roots of the algebraic equation
\begin{eqnarray}
&D^3\lambda^5_i-3D^2F\lambda^4_i+D(3F^2-2D\nu-h^2_i)
\lambda^3_i \nonumber \\
&+F(4D\nu-F^2+h^2_i)\lambda^2_i+  {\nu}(D\nu-2F^2+2qh^2_i)
\lambda_i \nonumber \\
& -\nu^2F=0\quad . \label{algebraic}
\end{eqnarray}
Eleven conditions for the ten constants of integration of
Eq.~(\ref{solution}) and for the probability current $J$ can be
determined at the points of discontinuity, by requiring
continuity, periodicity, and normalization of $\bf{P^s_i(x)}$.
This procedure leads to an inhomogeneous set of eleven linear
algebraic equations. Now, an exact formula for the current $J$ can
be obtained as a quotient of two determinants of the eleventh
degree. The exact formula, being complex and cumbersome, will not
be presented here, however, it will be used to find (i) the
dependence of the current $J$ on the tilting force $F$ and the
dependence of the mobility $m=J/F$ on the flatness
$\varphi=1/(2q)$, which are displayed in Figs.~2 and 3,
respectively, and (ii) the asymptotic limits of the current $J$
at low temperature and small external force.

Figure 2 shows the induced current $J$ as a function of the
external force $F$ for two different values of temperature and for
three different values of $d$ with fixed $\varphi=2.5$ and
$\nu=8$. In this figure, one also observes the hypersensitive
response at very low forcing, which apparently gets more and more
pronounced as the thermal noise strength $D$ decreases. For the
case $D=0,d=0.5$, the results of Monte Carlo simulations of the
current $J=J(F)$ are also presented. The tendency apparent in
Fig.~2, namely a decrease in the mobility for very low forcing as
the asymmetry of the potential grows, is also valid for large
asymmetries, e.g., when $d<0.05$.

To obtain more insight, we shall now study some asymptotic limits
of the current.

At the fast-noise limit, we allow $\nu$ to become large, holding
all the other parameters fixed. Thus,
 at very high frequencies of colored fluctuations,
 the system is under the influence of the average
 fluctuating potential. In the $\nu\to\infty$ limit,
 the current is then given by
\[
J=F+O(\nu^{-{1\over2}})\quad.
\]
The form of the leading term of the current $J$ is not confined to
the fast-noise limit. It is also valid for the asymptotic limit of
a high temperature, $D>>1$, and in the case of a large "load"
force $F$ ($F\to\infty$, all the other parameters fixed).

At the long-correlation-time limit $\nu\to0$, the equations
(\ref{master}) for $P^s_1(x),\:P^s_2(x)$, and $P^s_3(x)$ are
decoupled and the total current is given by the average of each
current for the corresponding potential configurations. In the
case of the symmetric potential, $d=1/2$, the current $J$
saturates at the value
\begin{eqnarray}
J &=& (1-2q)F \nonumber \\   &+& \frac{2q(4-F^2)^2 \:
\textrm{sh~}\!{(F/2D)}}{16D[\: \textrm{ch~}\!{\frac{1}{D}}-\:
\textrm{ch~}\!{\frac{F}{2D}}]-F(4-F^2) \:
\textrm{sh~}\!{\frac{F}{2D}}}\quad. \nonumber
\end{eqnarray}
For $F<2$, we can see that the current $J$ tends to $(1-2q)F$ as
$D\to0$. This result is consistent with the physical intuition
that the probability densities $P^s_1(x)$ and $P^s_3(x)$ are
$\delta$-distributed at deterministic stationary states (minimums
of potentials): the random variable $Z$ takes values $\pm1$ for a
sufficiently long time to allow the deterministic stationary
state to be formed.

In the case of zero temperature, $D=0$, and symmetric potential,
$d=1/2$, one finds from the exact formula that on the assumption
$F<2$ the current equals
\begin{equation}
J={\nu}F\frac{A_1C_2-C_1A_2}{B_1C_2-B_2C_1}\quad , \label{zero
temperature}
\end{equation}
where $A_i=F\{\alpha_i[F-(4-F^2)\eta_i]-2(1-2q)\}$,
$B_i=(\nu+16q)A_i+32q(1-2q)(2\alpha_i+F)$,
$C_i=qA_i+2(1-2q)[4\eta_i+Fq+2\alpha_i(1+F\eta_i)]$,
$\eta_i=F^{-1}(4-F^2)^{-1}\big[ F^2-4q-2\varepsilon_i
(4q^2+F^2(1-2q))^{1/2}]$, $\alpha_i=\textrm{th}(\nu\eta_i/4)$,
$i=1,2$, $\varepsilon_1=1$, $\varepsilon_2=-1$.

Thus, at the low-force limit, $F\to0$, the current will saturate
at the finite value
\begin{equation}
\lim_{F\to0}J=J_a=\frac{32{\nu}q(1-2q)}{(\nu+8)^2}\quad .
\label{low-force}
\end{equation}
As $J(F=0)=0$, the hypersensitive response is extremely
pronounced in this case, with the current picking up with an
infinite derivative at $F=0$ (see also Fig.~2). The asymptotic
current $J_a$ exhibits a bell-shaped (resonance) form as $\nu$ or
$q$ are varied. The optimal correlation time $\tau_m$ that
maximizes the current equals $1/8$, and the optimal flatness
$\varphi_m=1/(2q_m)=2$. It is remarkable that in the case of a
dichotomous noise $q=1/2$, the hypersensitive response disappears
and in the low forcing limit the leading-order term of the
current is proportional to
$F:\:J\approx{{\nu}F(\nu+12)/(\nu+8)^2}$.
\begin{figure} \centerline{
\psfig{file=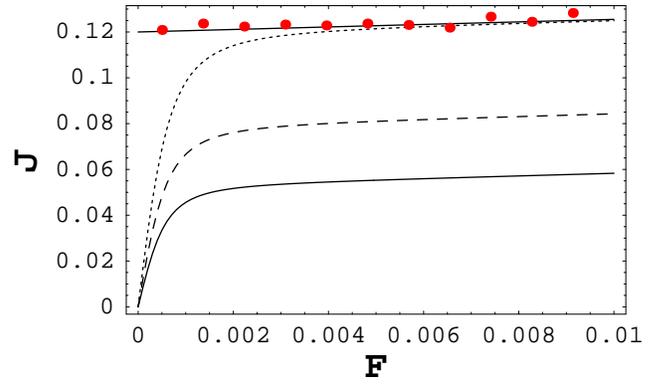,width=85mm}}\caption{ The current $J$ vs
applied force $F$ in the region of the hypersensitive response.
The flatness parameter equals $\varphi=2.5$ and the switching rate
$\nu=8$. Solid straight line: $D=0$, $d=0.5$. Dotted line:
$D=4\times10^{-8}$, $d=0.5$. Dashed line: $D=4\times10^{-8}$,
$d=0.2$. Solid curved line: $D=4\times10^{-8}$, $d=0.05$. The
filled dots on the solid straight line are obtained by means of
Monte Carlo simulations. Notice the jump of the current from the
zero level to the solid line corresponding to the infinite
derivative of $J(F)$ at $F=0$. }
\end{figure}

At the low forcing limit, $F\ll1$, a natural way to investigate
the behaviour of $J$ is to apply small-$F$ perturbation
expansions. A stationary solution of Eqs.~(\ref{solution}) and
(\ref{algebraic}) with $D\ne0$, $d=1/2$ is constructed in terms
of integer powers of $F$. The current can be expressed as
$J=Fm_1+F^2m_2+\dots$. We shall calculate the leading term of the
current $Fm_1$. Notably, the analysis of this section is valid for
the values of parameters satisfying the condition
$F<(2q{\nu}D)^{1/2}$. This condition results from the assumption
that the higher-order terms in the expansion of the roots of
Eq.~(\ref{algebraic}) are asymptotically smaller than the
lower-order terms held in the calculation. At sufficiently small
temperature, $D\ll\min\{1,2q\nu,8q/\nu\}$, the formula for the
leading-order term $Fm_1$ of the current is
\begin{equation}
J\approx{Fm_1}=\frac{8(1-2q)F}{(\nu+8)^2}
\sqrt{\frac{2q\nu}{D}}+FG\quad . \label{leading-order}
\end{equation}
Here the symbol $G$ stands for the terms which do not increase as
$D\to0$. An extreme sensitivity of the mobility $m$ to thermal
noise can be seen from the factor $D^{-1/2}$ in
Eq.~(\ref{leading-order}) that increases unboundedly as $D\to0$.
It can be seen easily that the functional dependence of the
mobility on the flatness $\varphi$ and on the correlation time
$\tau_c$ is of a bell-shaped form. The mobility $m_1$ reaches a
maximum at the flatness $\varphi_m=3$ and at the correlation time
$\tau_m=3/8$. The dependence of the mobility $m=J/F$ on the
parameters $q$ and $\nu$ for a fixed force value $F=10^{-5}$ and
for a fixed temperature $D=4\times10^{-8}$ is shown in Fig.~3. We
can see that the asymptotic formula (\ref{leading-order}) is in
excellent agreement with the exact results.
\begin{figure} \centerline{\psfig{file=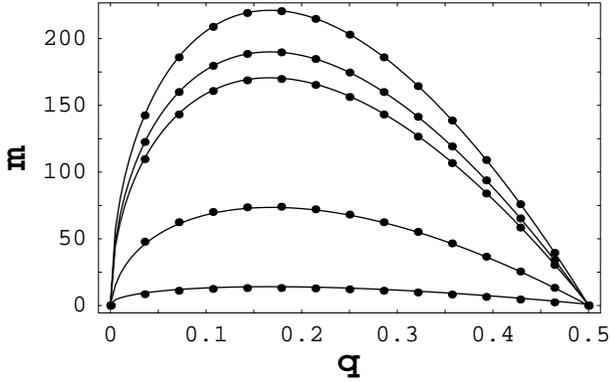,width=82mm}}
\caption{The mobility $m=J/F$ vs the flatness parameter
$q=1/(2\varphi)$ at $d=1/2$,~$D=4\times10^{-8}$, and $F=10^{-5}$.
The curves computed from the exact formula for the current $J$
correspond to the values of $\nu=8/3$,~$\nu=1$,~$\nu=8$,~$
\nu=0.1$,~$\nu=100$ from top to bottom. The non-monotonic sequence
of the values of $\nu$ stems from the bell-shaped dependence
$J=J(\nu)$.  Note that the maximum of the mobility lies at $q=1/6$
and $\nu=8/3$. The dots were computed by means of the asymptotic
formula ~(\ref{leading-order}).}
\end{figure}

Let us note that the sufficient condition, $F<\sqrt{2q{\nu}D}$,
has a distinct physical meaning: the characteristic distance of
thermal diffusion $\sqrt{D/\nu}$ is larger than the typical
distance $F/\nu$ for the particle driven by the deterministic
force $F$ in the state $z=0$ of the trichotomous noise. Let us
look at the latter statement more closely on the assumption that
$\nu\ll1$. For this assumption within the interval $(0,1)$, the
probability distributions $P_n^s(x)$, $n=1,3$, are, evidently,
concentrated at $x=0$ (or $x=1/2$). Next, we shall consider the
trajectory ($1:2:3:4:6$) in Fig.~1. The particles locked at the
potential minimum $1$ ($x=d=1/2$) will go at the initial time
$t=0$ to point $2$, where $z=0$. The first time when the noise
turns to either $z=1$ or $z=-1$ is denoted by $t_0$. As the time
of movement from $4$ to $6$ is much less than $t_0$, it is easy to
find that during the time interval $(0,t_0)$ the center of mass
has shifted by
\begin{displaymath}
\bigtriangleup{x}\approx\frac{1}{2
\sqrt{{\pi}Dt_0}}
\int_0^{Ft_0}\textrm{exp}\biggl\{-\frac{(x-Ft_0)^2}
{4Dt_0}\biggr\}dx\approx\frac{F\sqrt{t_0}}{2\sqrt{D\pi}}\quad .
\end{displaymath}

In the case of a trichotomous noise the probability $W(t)$ that
in a certain time interval $(0,t)$ the transitions
$z=0{\to}z=\pm1$ do not occur, is given by
$W(t)=\exp(-2q{\nu}t)$. The probability that the transition
$z\!=\!0\,{\to}\,z\!=\!-1$ occurs within the time interval
$(t,t+dt)$ is $q{\nu}dt$. Consequently,
\begin{displaymath}
\langle\bigtriangleup{x}\rangle={q\nu}\int_0^
\infty\textrm{e}^{-2q{\nu}t_0}\bigtriangleup{x} \: dt_0
\approx\frac{F}{8\sqrt{2q{\nu}D}}\quad .
\end{displaymath}
Considering that the average number of transitions per unit of
time into the state $z=0$ is $2q{\nu}(1-2q)$, we obtain
$J=2q\nu(1-2q)\langle\bigtriangleup{x}\rangle{
\approx}F(1-2q)\sqrt{2q\nu}/8\sqrt{D}$. Thus, we have obtained an
earlier result, namely,  Eq.~(\ref{leading-order}) for
${\nu}\ll1$. Formula (\ref{leading-order}) is one of our main
results. Note that the above procedure can be repeated in a
straightforward but tedious way for more complicated cases
involving asymmetric potentials and potentials with several
extrema per period. The phenomen is robust enough to survive a
modification of the multiplicative noise. The key-factor is the
noise flatness, indicating how long the noise level dwells on the
state $z=0$. If the flatness parameter is greater than $1$, the
effect does exist. For example, the multiplicative noise can also
be a Gaussian stationary process.

It is quite remarkable that the above results are also applicable
for amplifying adiabatic time-dependent signals, i.e., signals of
much greater periods than the characteristic time of establishing
a stationary distribution, even in the case of a small input
signal-to-noise ratio $F/\sqrt{D}\ll1$.

We emphasize that our mechanism of hypersensitive transport is of
a qualitatively different nature from a recently found effect,
where a noise-induced enhancement of the current of Brownian
particles in a tilted ratchet system has also been established
\cite{gin01,ben}. In the latter scenario, a system with a periodic
smooth potential exhibits hypersensitivity under the effect of
multiplicative dichotomous noise  because of noise-induced escape
through fixed points of the dynamics. This occurs because the
stable and unstable fixed points of the alternative dynamics,
which coincide in the absence of the tilt $F$, are shifted apart
by a small force (see also Ref. \cite{berd}). In the mechanism
reported here, we have a sharp periodic potential: the stable and
unstable fixed points of the dynamics coincide also for any small
tilt. The crossing of the location of the fixed points is achieved
by a combined influence of the flatness of the multiplicative
noise and a small tilt-forcing.

In a general case, if the potential is smooth and the flatness of
multiplicative noise is greater than $1$, both mechanisms play an
important role and should be taken into account. Our calculations
show that the factor $F\sqrt{\nu/D}$ in Eq.~(\ref{leading-order})
is generated by thermal diffusion in the state $z=0$, while the
circumstance that the potential is sharp has no effect on this
factor. On the other hand, for adiabatic switching, the mechanism
described in Ref. \cite{gin01} generates the current
$J\sim{\nu}F/\sqrt{D}$. Consequently, our mechanism for
sufficiently small switching rates induces hypersensitive
transport more effectively than the one proposed by Ginzburg and
Pustovoit. This conclusion is in agreement with the results of
Ref.\cite{gin01}, presenting numerical simulations of the
phenomenon of hypersensitive transport based on a phase model with
the multiplicative colored Gaussian noise ($\varphi=3$). It is
established that in the case of low switching rates the transport
for the Gaussian noise appears to be more effective than for
dichotomous stimuli. Regrettably, the authors of Ref.\cite{gin01}
did not consider the role of noise flatness and the physics of
this discrepancy.

In conclusion, the reported mechanism of generating hypersensitive
transport by the flatness of multiplicative noise is of general
relevance for many physical, biological, and chemical systems, and
may provide another possibility to control signal amplification.
The sensitivity of system response to small input signals can be
either enhanced or suppressed by changing the noise parameters
(flatness, correlation time, temperature). In agreement with
Ref.\cite{ger01}, we believe that the phenomenon proposed may also
shed some light on the ability of biological systems to detect
weak signals in a noisy environment.

We acknowledge partial support by the Estonian Science Foundation
Grant Nos.~4042 and 5662 and by the International Atomic Energy
Agency Grant No.12062.

\end{document}